\documentclass[%
 reprint,
 amsmath,amssymb,
 aps
]{revtex4-1}

\usepackage{graphicx}
\usepackage{dcolumn}
\usepackage{bm}

\begin{document}

\title{Topological phases in the non-Hermitian Su-Schrieffer-Heeger model}
\author{Simon Lieu}
\email{simonklieu@gmail.com}
\affiliation{
   Blackett Laboratory, Imperial College London, London SW7 2AZ, United Kingdom
   }

\date{\today}
\begin{abstract}

We address the conditions required for a $\mathbb{Z}$ topological classification in the most general form of the non-Hermitian Su-Schrieffer-Heeger (SSH) model. Any chirally-symmetric SSH model will possess  a ``conjugated-pseudo-Hermiticity" which we show is responsible for a quantized ``complex" Berry phase. Consequently, we provide the first example where the complex Berry phase of a band is used as a quantized invariant to predict the existence of \textit{gapless} edge modes in a non-Hermitian model. The chirally-broken, $PT$-symmetric model is studied;  we suggest an explanation for why the topological invariant is a global property of the Hamiltonian. A geometrical picture is provided by examining eigenvector evolution on the Bloch sphere. We justify our analysis numerically and discuss relevant applications.
\end{abstract}

\maketitle

\section{Introduction}

Recent studies have suggested that non-Hermitian analogs of the Su-Schrieffer-Heeger (SSH) model are relevant in describing one-dimensional (1D) topological behavior in systems with gain and/or loss. Specifically, a number of theoretical works have suggested methods to achieve $PT$-symmetric versions of the SSH model in an optical setting \cite{Schomerus1, Rudner1,Esaki1, Yuce1, Zhu1}, with recent experimental successes demonstrating the existence of robust edge states \cite{Poli1,Rudner2,Weimann1,Xue1}. Thus an active area of research aims to characterize topological phenomena in non-Hermitian models \cite{Fu1, Hughes1,Ghosh1,Esaki1,Liang1,Leykam1,Klett1}.

A crucial assumption of the ten-fold way classification of topological insulators is Hermiticity of the Hamiltonian \cite{Ryu1}. In 1D Hermitian models, the presence of either chiral \cite{Ryu1} or inversion symmetry \cite{Lu1} is responsible for $\mathbb{Z}$ topological phases.  What symmetries are necessary to ensure a $\mathbb{Z}$ classification in a 1D non-Hermitian model?

We frame our analysis in terms of the non-Hermitian generalization of the Berry phase: the ``complex" Berry (cBerry) phase \cite{Berry1,Garrison1, Dattoli1, Mostafazadeh1}. In doing so, we find that any chirally-symmetric SSH model will possess a $\mathbb{Z}$ invariant, including systems where inversion symmetry is broken. Fundamentally this is because any chiral model will possess a ``conjugated-pseudo-Hermiticity" which is responsible for a quantized cBerry phase. We also discuss the chirally-broken, $PT$-symmetric SSH model by considering an imaginary staggered potential \cite{Schomerus1}. We argue that because bulk band crossings occur away from the topological transition point, the topological invariant must be a global property of the Hamiltonian.

A central goal of this work is to clarify the role of the complex Berry phase in non-Hermitian models. In the absence of Hermiticity, it is unclear what quantity serves as the bulk topological index in general. Previous studies have pointed to quantization of the real Berry phase \cite{Schomerus1,Ling1,Weimann1}, global cBerry phase \cite{Liang1}, Pfaffian \cite{Menke1}, and various winding numbers \cite{Rudner1, Esaki1,Leykam1} to characterize a transition. In this study, we highlight the importance of the complex Berry phase of a band by demonstrating that its quantization will accurately predict the existence of gapless edge modes in non-Hermitian models. We posit that if the system has gapped edge modes, then the topological index must be associated with the entire Hamiltonian, not a specific band. This is in agreement with previous studies \cite{Liang1,Esaki1,Rudner1}.

While a number of studies have focused on the $PT$-symmetric version of the model, the non-Hermitian, chirally-symmetric case has received less attention. Such a model arises when discussing plasmonic dispersion on a bipartite lattice \cite{Pocock1, Ling2} and in optical analogs of the Hatano-Nelson model \cite{Hatano1,Longhi2,Longhi3,Longhi4}. We speak more concretely concerning problems which benefit from our analysis towards the end of the work.

\section{Model and Results}

We begin by studying the most general 1D nearest-neighbor tight binding model on a bipartite lattice without assuming Hermiticity. Consider $N$ pairs of particles on a finite chain with Hamiltonian
\begin{multline}
H^{\text{hop}}=v_{1}\sum_{n=1}^{N}\left|n,B\right\rangle \left\langle n,A\right|+v_{2}\sum_{n=1}^{N}\left|n,A\right\rangle \left\langle n,B\right| \\ +w_{1}\sum_{n=1}^{N-1}\left|n+1,A\right\rangle \left\langle n,B\right|+w_{2}\sum_{n=1}^{N-1}\left|n,B\right\rangle \left\langle n+1,A\right|\label{eq:ham}
\end{multline}
where we do \textit{not} assume $v_{1}=v_{2}^{*},w_{1}=w_{2}^{*}$ which defines the Hermitian case \cite{SSH1,Asboth1}. In addition to the hopping, we will consider an imaginary staggered potential when discussing the $PT$-symmetric case, given by the term
\begin{equation}
H^{\text{pot}}=iu\sum_{n=1}^{N}\left(\left|n,A\right\rangle \left\langle n,A\right|-\left|n,B\right\rangle \left\langle n,B\right|\right).
\end{equation}
The full Hamiltonian reads $H=H^{\text{hop}}+H^{\text{pot}}$. If periodic boundary conditions are assumed, then the system possesses discrete translational invariance and hence may be diagonalized according to Bloch's theorem by considering eigenvectors of the form
\begin{equation}
\left|\psi_{k}\right\rangle =\frac{1}{\sqrt{N}}\sum_{n=1}^{N}e^{ikn}\left(a_{k},b_{k}\right)\left(\left|n,A\right\rangle ,\left|n,B\right\rangle \right)^{T}. \label{eq:blochstates}
\end{equation}
Substituting this ansatz into (\ref{eq:ham}) leads to the equation
\begin{equation}
\left(\begin{array}{cc}
iu & w_{1}e^{-ik}+v_{2}\\
w_{2}e^{ik}+v_{1} & -iu
\end{array}\right)\left(\begin{array}{c}
a_{k}\\
b_{k}
\end{array}\right)=E(k)\left(\begin{array}{c}
a_{k}\\
b_{k}
\end{array}\right)\label{eq:bulkH}
\end{equation}
where the $2\times 2$ matrix above is defined as the bulk Hamiltonian $H(k)$. Solving for the bulk dispersion $E(k)$ and corresponding eigenvectors provides us with all the modes in the model except edge modes, which only appear in the finite chain Hamiltonian without periodic boundary conditions (\ref{eq:ham}). The bulk-boundary correspondence states that an invariant calculated from the bulk Hamiltonian can predict how many gapless, topologically-protected edge modes to expect in the finite system \cite{Franz1}.

\begin{figure}
\begin{centering}
\includegraphics[scale=0.8]{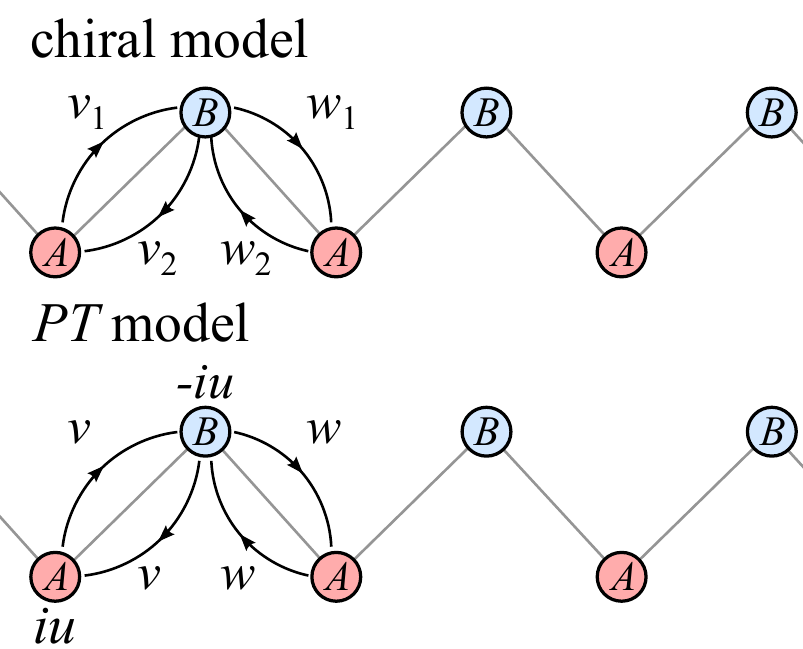}
\par\end{centering}
\caption{\label{fig:chain} \textit{Top}: Chiral SSH with non-Hermitian hopping parameters, no potential. \textit{Bottom}: Chirally-broken, $PT$-symmetric SSH with non-Hermitian staggered potential.}
\end{figure}

In what follows, we will be concerned with two distinct scenarios: 1) completely general hopping in the absence of a staggered potential $u=0$, which is referred to as the chirally-symmetric case; 2) real, symmetric hopping (up to a gauge) in the presence of an imaginary potential $v_{1}=v_{2}=v\in\mathbb{R},w_{1}=w_{2}=w\in\mathbb{R},u\neq0,u\in \mathbb{R}$, which is referred to as the $PT$-symmetric case. Figure \ref{fig:chain} gives an illustration of the two set-ups.

We discuss the symmetries in the absence of a staggered potential $u=0$. Restricting hopping to nearest-neighbors  implies that the Hamiltonian possesses chiral symmetry, defined by $\sigma_{z}H(k)\sigma_{z}=-H(k)$, where $\sigma_i$ refers to Pauli matrices. This results in eigenvalues $E(k)$ which come in $\pm$ pairs at a given $k$. Chiral symmetry implies a conjugated-pseudo-Hermiticity $H^{\dagger}(k)=\sigma_{x}H(k)^{*}\sigma_{x}$ since both conditions are satisfied only if the bulk Hamiltonian lacks a term proportional to $\sigma_z$. We will show that this is the necessary ingredient for a quantized cBerry phase. If $w_{1}=w_{2},v_{1}=v_{2}$ then $H$ possesses inversion symmetry defined as $\sigma_{x}H(k)\sigma_{x}=H(-k)$, which implies $E(k)=E(-k)$.

In the presence of an imaginary staggered potential $u\neq0$ and real hopping $v_{1}=v_{2}=v,w_{1}=w_{2}=w$, chiral symmetry is broken but the system possesses $PT$-symmetry defined by $\sigma_{x}H(k)\sigma_{x}=H(k)^{*}$. Generally, $PT$-symmetric models have two parameter regimes called the ``broken" and ``unbroken" phases \cite{Bender1,Bender2}. The unbroken phase has attracted much attention, since it is defined by a fully real spectrum and eigenvectors which are $PT$-symmetric  $\left|\psi_{k}\right\rangle =\sigma_{x}\left|\psi_{k}\right\rangle ^{*}$. The unbroken phase for our model occurs when $u<|v-w|$. The $PT$-symmetric model possesses a pseudo-anti-Hermiticity $H(k)^{\dagger}=-\sigma_{z}H(k)\sigma_{z}$, which is responsible for a topological transition (discussed below). Note that pseudo-anti-Hermiticity reduces to chiral symmetry in the case of a Hermitian Hamiltonian.

Chiral symmetry ensures that $N_A-N_B$ is a topological invariant, where $N_{A/B}$ is the number of zero-energy, left edge modes with support on sublattice $A/B$ \cite{Ryu2,Asboth1}. This is because topologically-protected edge modes are their own chirally-symmetric partners, which pins their energy to zero. By considering the two completely dimerized regimes $|v_{1,2}|=0,|w_{1,2}|\neq0$;$|v_{1,2}|\neq0,|w_{1,2}|=0$, we find that this invariant changes from one to zero and hence infer that a topological transition must occur somewhere in between these two limits. Similarly, Esaki \textit{et al.} argue that pseudo-anti-Hermiticity acts analogously to chiral symmetry with the caveat that edge modes must appear gapless in the real plane only \cite{Esaki1}. What bulk invariant is responsible for these edge states?

Before specializing to the model (\ref{eq:bulkH}), we introduce the Berry phase for non-Hermitian Hamiltonians. The ``complex'' Berry phase for band $n$ is defined by
\begin{equation}
Q_{n}^{c}=i\int_{BZ} \left\langle \lambda_{k}^{n}\left|\frac{\partial}{\partial k}\right|\psi_{k}^{n}\right\rangle dk\label{eq:berry}
\end{equation}
where $n$ labels the band index, $\left|\psi\right\rangle ,\left|\lambda\right\rangle $ are eigenvectors of $H,H^{\dagger}$ respectively (namely the right and left eigenvectors of the Hamiltonian), and the integral is taken over the 1D Brillouin zone \cite{Garrison1, Dattoli1, Mostafazadeh1}. A generalization of Berry's original argument \cite{Berry1} suggests that $Q_{n}^{c}$ is the geometrical phase picked up by an adiabatic deformation across the Brillouin zone, which arises fundamentally because the states $\left|\psi^{n}\right\rangle $ no longer form an orthogonal basis while bi-orthonormality constraints are satisfied $\left\langle \lambda^{n}\left|\psi^{m}\right.\right\rangle =\delta_{nm}$.

We briefly mention that in the literature, some studies have suggested that the quantization of the ``real'' Berry phase, defined by
\begin{equation}
Q_{n}^{r}=i\int_{BZ}  \left\langle \psi_{k}^{n}\left|\frac{\partial}{\partial k}\right|\psi_{k}^{n}\right\rangle  dk\label{eq:rBerry}
\end{equation}
where $\left\langle \psi^{n}\left|\psi^{n}\right.\right\rangle =1$,  implies a topological classification in non-Hermitian systems \cite{Schomerus1,Weimann1,Ling1}. We will find instances where $Q_{n}^{r}$ does \textit{not} predict the existence of gapless edge modes, while its complex counterpart does in all situations encountered.

We now study the chirally-symmetric model, \textit{i.e.} $u=0$. The right/left eigenvectors are related via conjugated-pseudo-Hermiticity by $\left|\lambda_{k}^{\pm}\right\rangle =\sigma_{x}\left|\psi_{k}^{\pm}\right\rangle ^{*}$. Additionally the chiral symmetry relates the positive/negative bands by $\left|\psi_{k}^{\pm}\right\rangle =\sigma_{z}\left|\psi_{k}^{\mp}\right\rangle $. Combining these together leads to a useful parametrization
\begin{equation}
\begin{aligned}
\left|\psi_{k}^{\pm}\right\rangle &=\frac{1}{\sqrt{2\sin\theta_{k}\cos\theta_{k}}}\left(\begin{array}{c}
e^{-i\phi_{k}}\cos\theta_{k}\\
\pm\sin\theta_{k}
\end{array}\right) \\ \left|\lambda_{k}^{\pm}\right\rangle &=\frac{1}{\sqrt{2\sin\theta_{k}\cos\theta_{k}}}\left(\begin{array}{c}
e^{-i\phi_{k}}\sin\theta_{k}\\
\pm\cos\theta_{k}
\end{array}\right)\label{eq:leftright}
\end{aligned}
\end{equation}
where eigenvectors are normalized to obey the conditions $\left\langle \lambda_{k}^{\pm}\left|\psi_{k}^{\pm}\right.\right\rangle =\delta_{++;--}$,
$\left\langle \psi_{k}^{\pm}\left|\psi_{k}^{\pm}\right.\right\rangle \neq1,0$.
Substituting (\ref{eq:leftright}) into the expression for $Q^{c}$ in (\ref{eq:berry}), after some algebra, we find
\begin{equation}
Q_{\pm}^{c}=\frac{1}{2}\int_{BZ}\dot{\phi}_{k}dk=\frac{1}{2}\left(\left.\phi_{k}\right|_{k=-\pi}^{k=\pi}\right)
\end{equation}
which is clearly quantized to be an integer multiple of $\pi$. Note that our parametrization (\ref{eq:leftright}) is ill-defined in the case when $\theta_{k}=n\pi/2,n\in\mathbb{Z}$ due to the diverging normalization constant. However $\theta_{k}$ attains such values only when a band gap closes for the chiral system. This is because chiral symmetry precludes a $\sigma_z$ term in $H(k)$, hence the eigenvectors $\left|\psi_{k}\right\rangle =(1,0)^T,(0,1)^T$ can only occur if one of the off-diagonal elements in the bulk Hamiltonian is zero, and hence at a band crossing. (This point is also known as an exceptional point \cite{Heiss1}.) Thus the cBerry  phase is a topological invariant: as long as adiabatic deformations to the Hamiltonian preserve the band gap, the cBerry phase will be integer quantized in units of $\pi$.

In Appendix \ref{sec:apA} we show that the real Berry phase in a chiral model is not quantized if inversion symmetry is broken. Note that in the cBerry phase analysis above there is no reference to inversion symmetry and relationships between right/left eigenvectors arise due to chiral symmetry alone. Later we will confirm numerically that there are chiral models where inversion symmetry is broken which possess gapless edge modes. Interestingly, recent studies have found that the non-Hermitian 2D Chern number is not sensitive to the combination of right/left eigenvectors used in the expression for the Berry curvature, in contrast to the 1D Berry phase \cite{Fu1}.

In the previous section we have argued that if the system has chiral symmetry, then generically the cBerry phase will be quantized to an integer value away from a band crossing. It is well-known that a real, staggered potential will destroy quantization in the Hermitian SSH since such a term breaks both chirality and inversion \cite{Lu1, Ryu1}. Physically this is because edge states acquire an energy splitting and hence can be adiabatically removed without closing a band gap. In the $PT$-symmetric model, edge state are gapless in the real plane but acquire an imaginary energy gap according to $E_{\text{edge}}=\pm ui$. However, due to pseudo-anti-Hermiticity arguments we know that a topological transition must take place. Liang and Huang  \cite{Liang1} derived the expression for the complex Berry curvature
\begin{equation}
\left\langle \lambda_{k}^{\pm}\left|\frac{\partial}{\partial k}\right|\psi_{k}^{\pm}\right\rangle^{PT} = \frac{1}{2} \left(1\pm\cos\xi_{k}\right)\dot{\eta}_{k}
\end{equation} 
where $\cos \xi_{k}=\frac{iu}{\sqrt{\left|v+we^{-ik}\right|^2-u^2}}$, $e^{-i\eta_{k}}=\frac{v+we^{-ik}}{\left|v+we^{-ik}\right|}$. Upon integration across the Brillouin zone, we find that the cBerry phase of a given band is not quantized (which is a common feature for gapped edge modes) but interestingly their sum is quantized.

\begin{figure}
\begin{centering}
\includegraphics[scale=0.7]{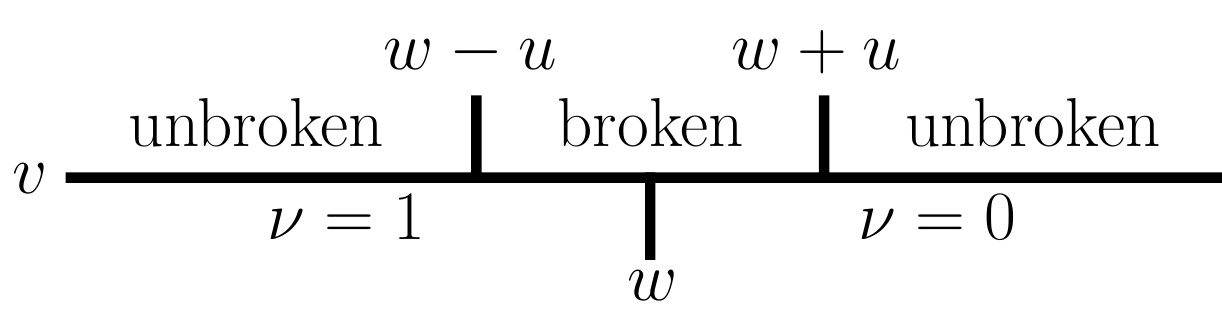}
\par\end{centering}
\caption{\label{fig:phaseDiag} One-dimensional phase diagram in $v$ for the $PT$-symmetric case $v_{1}=v_{2}=v\in\mathbb{R},w_{1}=w_{2}=w\in\mathbb{R}$ at a fixed value of $u,w>0$. The system moves in and out of the $PT$-broken phase as a function of $v$. $\nu$ is the topological index which predicts how many pairs of gapless-real-energy, edge modes exist in the system. The bulk spectrum is gapless in the entire $PT$-broken phase.}
\end{figure}

The $PT$-symmetric topological transition is fundamentally different from the normal ``topological insulator" paradigm. In the Hermitian SSH, edge modes are removed at a unique point where the bands cross; conversely, the entire $PT$-broken phase has bulk band crossings, yet edge modes still exist in the $PT$-broken region as long as $v<w$.  Indeed the transition point occurs when bulk modes are degenerate with edge modes, which are gapped in the imaginary plane. A phase diagram is given in Fig. \ref{fig:phaseDiag}.

Previous studies have suggested various global indices to characterize the number of pairs of gapless-real-energy edge modes in the $PT$-symmetric model. Liang and Huang \cite{Liang1} found that the global cBerry phase $Q^c_G=Q^c_+ + Q^c_-$  accurately captures the transition. Alternatively, Esaki \textit{et al.} \cite{Esaki1} define a winding number using the $Q$-matrix classification technique \cite{Ryu1}. It is not surprising that the topological invariant in the $PT$ model is not a property of the band, but rather of the entire Hamiltonian since bulk band crossings occur in the entire broken region and band indices lose their meaning at a band crossing.

We now provide a geometrical picture which further justifies our results for the chiral model by projecting eigenvectors onto the Bloch sphere. The most general 2D right-eigenvector can be parametrized as
\begin{equation}
\left|\psi'(\alpha_{k},\beta_{k})\right\rangle =\left(\begin{array}{c}
\cos(\beta_{k}/2)\\
e^{i\alpha_{k}} \sin(\beta_{k}/2)
\end{array}\right).
\end{equation}
For each eigenvector we may calculate the Bloch vector, defined as
\begin{equation}
\mathbf{b}_{k}=\left\langle \psi_{k}'\left|\boldsymbol{\sigma}\right|\psi_{k}'\right\rangle 
\end{equation}
at each $k$ point, where $\boldsymbol{\sigma}$ is the vector of Pauli matrices. $\alpha_k,\beta_k$ correspond to the azimuthal and polar angles of $\mathbf{b}_k$ respectively.

\begin{figure}
\begin{centering}
\includegraphics[scale=0.25]{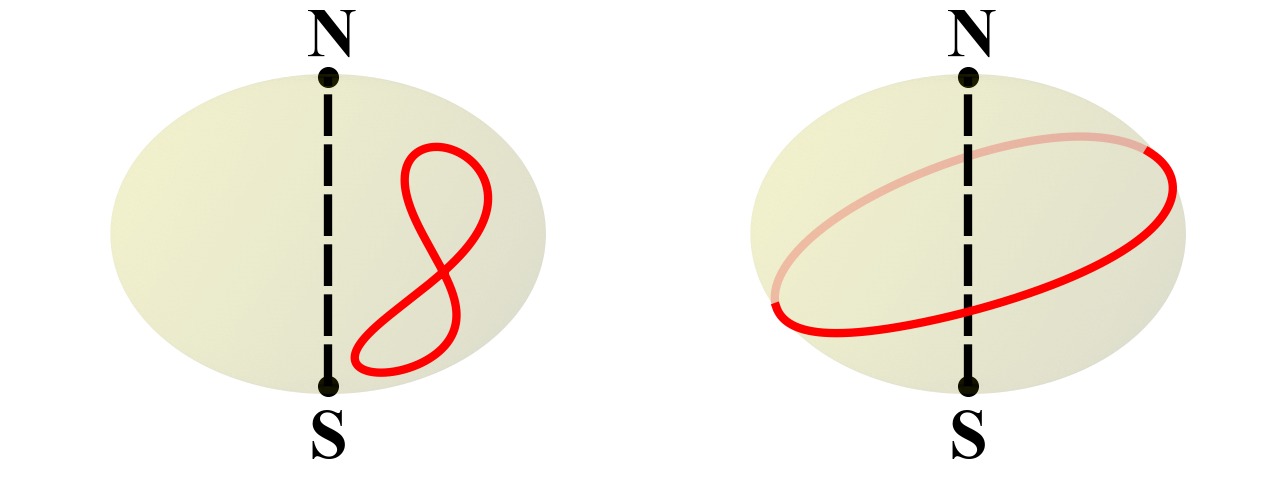}
\par\end{centering}
\caption{\label{fig:bloch}The lower-band vectors $\mathbf{b}_{k}$ (red) on the Bloch sphere in the range $k\in\left[-\pi,\pi\right)$ for a chiral model when: $Q^c_{\pm}=0$ (left); $Q^c_{\pm}=\pi $ (right). The north/south axis is dashed in black. Systems with distinct winding number cannot be deformed into each other without passing through the north or south pole (black dots), which can only occur at a band crossing in the chiral model. $\mathbf{b}_{k}$ are \textit{not} constrained to the equator.}
\end{figure}

In Fig. \ref{fig:bloch} we plot the eigenvector evolution on the Bloch sphere across the Brillouin zone for  two chiral systems with distinct cBerry phases. The topological classification arises due to the result from homotopy theory $\pi_{1}(S^{2} / \{n,s\})=\mathbb{Z};n=\left(0,0,1\right)^{T},s=\left(0,0,-1\right)^{T}$,  where we map the one-dimensional path across the Brillouin zone onto the surface of the Bloch sphere and observe that distinct winding numbers around the $z$-axis cannot be smoothly deformed into each other without passing the loop through the north or south pole. The north/south poles of the Bloch sphere are attained if $\left|\psi ' (k)\right\rangle =(1,0)^{T},(0,1)^{T}$ respectively, which (as explained previously) can only occur at a band crossing. If we restrict adiabatic deformations to those which preserve band openings then the poles are not accessible, hence distinct winding numbers are topologically preserved. The source of the $\mathbb{Z}$ classification in the non-Hermitian model is to be contrasted with the Hermitian SSH model, where Bloch vectors are constrained to lie on the equator, hence $\pi_{1}(U(1))=\mathbb{Z}$.

In the presence of a real, staggered potential in the Hermitian SSH, topological classification is lost due to the result $\pi_{1}(S^{2})=0$, since eigenvectors can reach the poles without closing a band gap. This simple one-band picture breaks down in the presence of an imaginary potential, \textit{i.e.} the $PT$-symmetric case, since band crossings occur in the broken phase. At the point of the band crossing, it is unclear how to continue the Bloch vector loop due to the degeneracy.

\begin{figure}
\begin{centering}
\includegraphics[scale=0.1]{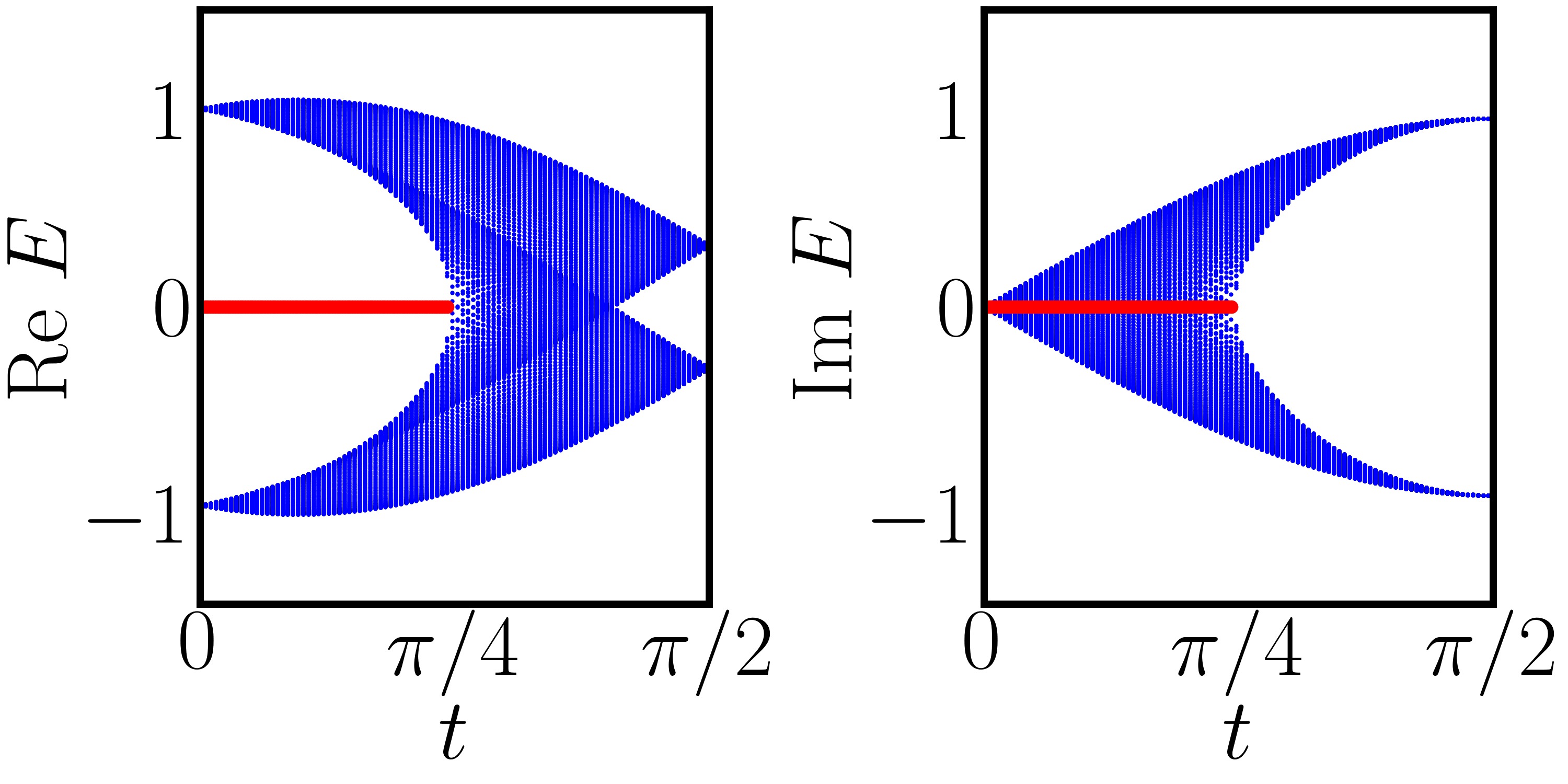}
\par\end{centering}
\caption{\label{fig:invSym}Chiral, inversion-symmetric: $u=0,v_{1}=v_{2}=\exp\left(i\pi/5\right)\sin t,w_{1}=w_{2}=\cos t,N=100$. Edge modes are in red; bulk modes are in blue.}
\end{figure}

\begin{figure}
\centering{}
\includegraphics[scale=0.1]{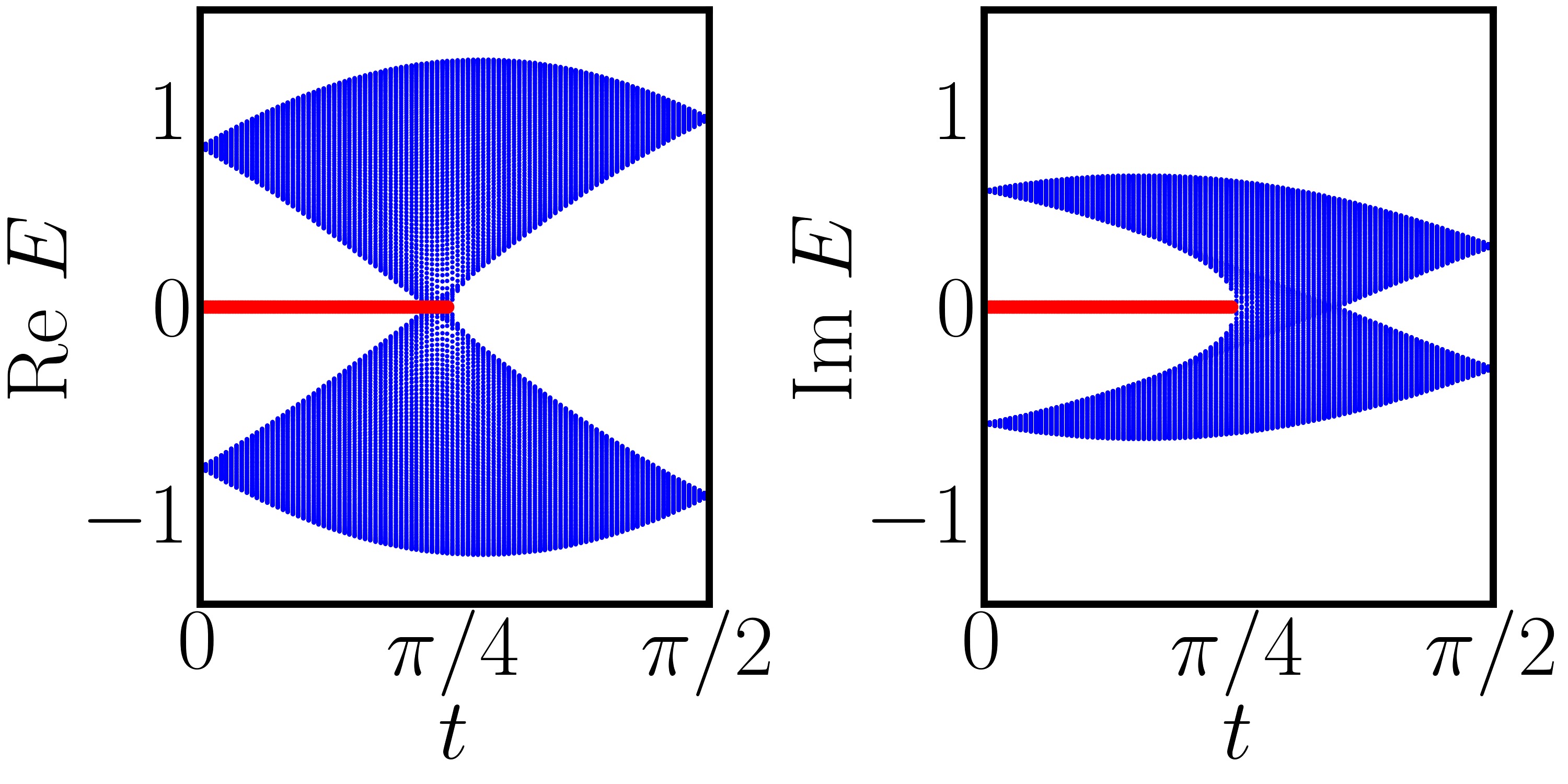}
\caption{\label{fig:invBreak}Chiral, inversion-breaking: $u=0,v_{1}=\sin t,v_{2}=\exp\left(i\pi/10\right)\sin t,w_{1}=\cos t,w_{2}=\exp\left(i\pi/5\right)\cos t,N=100$. Edge modes are in red; bulk modes are in blue.}
\end{figure}

\begin{figure}
\begin{centering}
\includegraphics[scale=0.1]{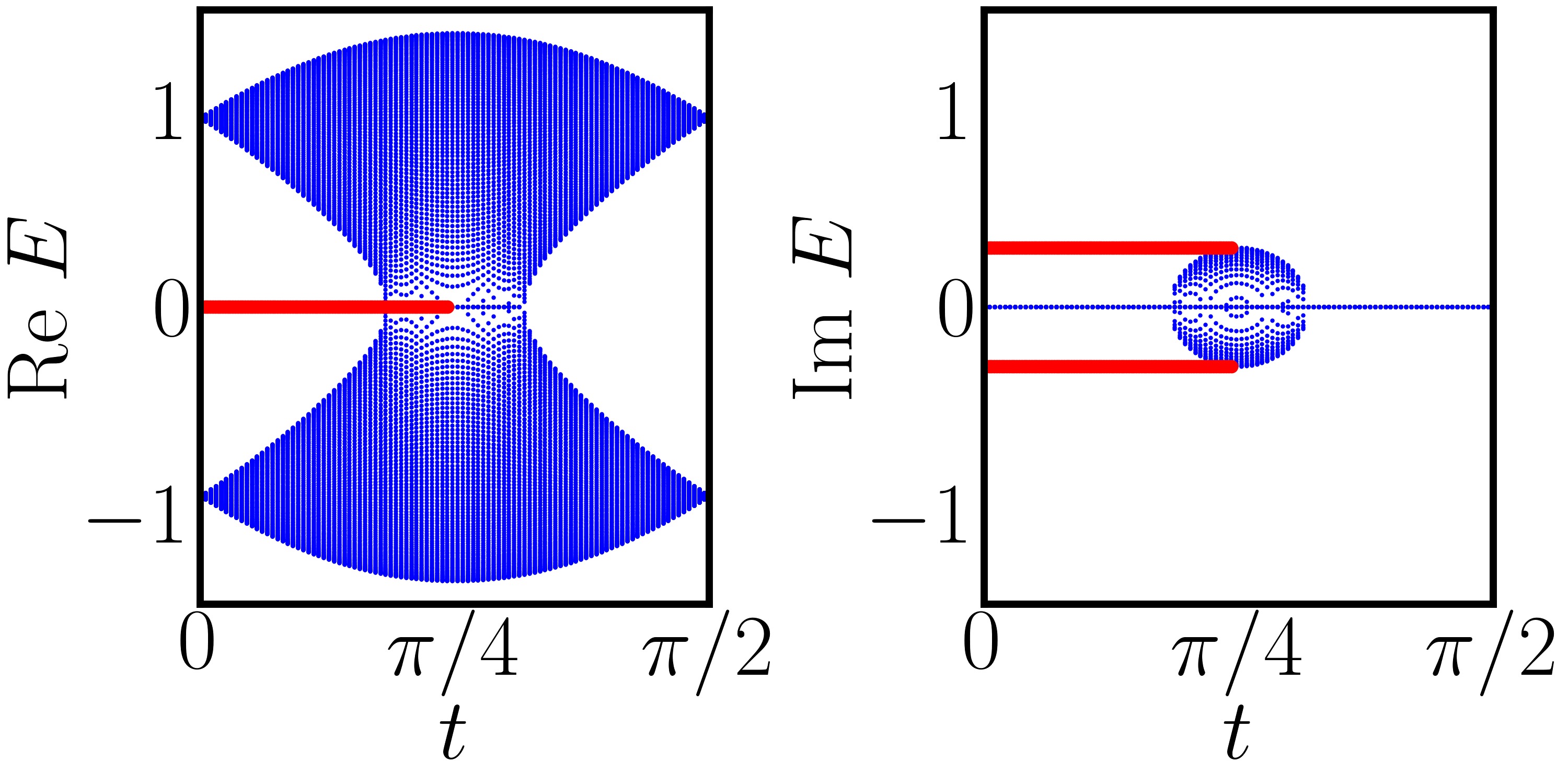}
\par\end{centering}
\caption{\label{fig:PT}$PT$-symmetric: $u=0.3,v_{1}=v_{2}=\sin t,w_{1}=w_{2}=\cos t,N=100$. Edge modes are in red; bulk modes are in blue.}
\end{figure}

We proceed to numerically verify our analysis.  In Figs. \ref{fig:invSym}, \ref{fig:invBreak}, and \ref{fig:PT}, the real and imaginary energy eigenvalues are plotted for the finite chain Hamiltonian (\ref{eq:ham}) for three distinct set-ups: 1) chiral, inversion-symmetric,  2) chiral, inversion-breaking, 3) $PT$-symmetric, with details found in the captions.

In the presence of chiral symmetry, \textit{i.e.} Figs. \ref{fig:invSym} and \ref{fig:invBreak},  the edge modes are gapless in both real and imaginary energy. Crucially, while bulk band gaps may close in the real or imaginary plane away from the topological transition point, the only time they close simultaneously is at the transition. In the inversion-broken scenario, the bulk spectrum is no longer invariant under the transformation $k \rightarrow -k$ (not shown), however we see that a topological transition persists. In Appendix \ref{sec:apB} we elaborate on the spectrum of the chiral, inversion-symmetric case.

In the $PT$-symmetric plot, Fig. \ref{fig:PT},  the two $PT$-unbroken phases (defined by purely real bulk modes) are separated by a $PT$-broken phase where some bulk modes become purely imaginary. Moreover, note that there are a pair of gapped edge modes in the imaginary plane when $v<w$. At the point of the topological transition, bulk modes become degenerate with edge modes. In the $PT$-broken phase the bulk spectrum is gapless.

\section{Applications}

While we have focused on a family of SSH models, our main results are expected to generalize to other 1D non-Hermitian systems. Specifically, the cBerry phase is expected to be quantized for any chiral model away from a bulk band crossing. We note that the 1D model in Ref. \cite{SchomerusPRL} falls outside the scope of this framework, since it possesses chiral symmetry yet has bulk band crossings away from the topological transition point, implying that the cBerry phase is ill-defined in these regions. It is not clear that the cBerry phase can be used as a topological invariant for this model.

The original motivation for this study began by considering the problem of plasmonic dispersion on bipartite chains of nanoparticles. The self-consistent Green's equation for this model is of the form $G\mathbf{p}=\alpha(\omega)^{-1}\mathbf{p}$ where $\mathbf{p}$ is the polarization vector of the chain, $\alpha(\omega)^{-1}$ is the inverse polarizability which can be mapped to the real and imaginary part of the plasmonic dispersion frequency, and $G$ is a Green's function matrix which defines how the electric field of a radiating nanoparticle effects neighboring polarization \cite{Ling2}. Once radiative and retardation affects are taken into account, the hopping elements respect inversion-symmetry, but acquire a Hemiticity-breaking phase \cite{Pocock1, Downing1,Ling2,Weber1}.

The $PT$-symmetric Hamiltonian studied in this work has been theoretically proposed in the context of complex photonic lattices with alternating gain/loss \cite{Schomerus1} and recently been realized in experiment with the observation of robust edge states \cite{Weimann1}. We note that $PT$-symmetric SSH models and 1D ``non-Hermitian quantum walks'' are closely related \cite{Rudner1,Liang1}. In the language of our analysis, the non-Hermitian quantum walk Hamiltonian is simply $H^{qw}=H^{PT}-iu\mathbb{I}$,  where $\mathbb{I}$ is the identity matrix. Recent optical experiments have indeed found edge states in the non-Hermitian quantum-walk model \cite{Rudner2,Xue1}.

From a numerical perspective, our study of the non-Hermitian, chiral model provides an exact expression for the cBerry phase that one may hope to match with computational methods \cite{Wagner1}.

\section{Summary}

In summary, we have investigated the topological properties of non-Hermitian SSH models with various symmetries, and in the process shed light on the role of the complex Berry phase. We have demonstrated that any chirally-symmetric SSH model will possess a quantized cBerry phase for each band which serves as a good topological invariant to predict the existence of gapless edge modes. We have provided an argument which explains why the $PT$-symmetric model possesses a global invariant: pseudo-anti-Hermiticity leaves edge modes gapless in real-energy, but gapped in imaginary-energy. This implies that bulk band crossings can occur away from the topological transition point which suggests that the invariant is a property of the entire Hamiltonian, not individual bands. From the viewpoint of classification theory, it appears that chiral symmetry and pseudo-anti-Hermiticity are the relevant symmetries necessary to categorize non-Hermitian 1D transitions, though future work needs to address this in a more systematic way \footnote{It is interesting to note that the $PT$ model falls into the AI class of the ten-fold way which is topologically trivial in 1D Hermitian models}. We have used homotopy arguments to justify the $\mathbb{Z}$ classification of the chiral model by projecting eigenvectors onto the Bloch sphere and evolving them across the Brillouin zone. We have verified our analysis numerically and discussed relevant applications of the study. Future work will aim to extend these ideas to higher dimensions.

\section*{Acknowledgements}

SL would like to sincerely thank Farzan Vafa, Caleb Q. Cook, and Derek K. K. Lee for useful discussion, and especially Simon R. Pocock for assistance with appendix calculations. SL is grateful for financial support from the Imperial College President's Scholarship.

\appendix

\section{\label{sec:apA} Real Berry phase in a chiral model}

There has been some debate concerning whether the real or complex Berry phase serves as the topological invariant for non-Hermitian models \cite{Weimann1,Ling1,Schomerus1,Liang1}. In the main text we have demonstrated that the complex Berry phase is quantized as long as the Hamiltonian has chiral symmetry. We now demonstrate that if inversion symmetry is broken, then the real Berry phase is not quantized for the chiral model. Numerically we found zero-energy edge states exist even in the absence of inversion symmetry, implying that the complex Berry phase serves as the proper index which predicts gapless modes in non-Hermitian models.

Consider the chiral model, \textit{i.e.} $u=0$
\begin{equation}
H\left(k,w_{1},w_{2},v_{1},v_{2}\right)=\left(\begin{array}{cc}
0 & w_{1}e^{-ik}+v_{2}\\
w_{2}e^{ik}+v_{1} & 0
\end{array}\right)\label{eq:bulkH-1}
\end{equation}
and note the following relation
\begin{equation}
H\left(k,w_{1},w_{2},v_{1},v_{2}\right)=\sigma_{x}H\left(-k,w_{2},w_{1},v_{2},v_{1}\right)\sigma_{x}. \label{eq:inv}
\end{equation}
This can be viewed as an ``inversion condition'' which states that if we view the system from the opposite end and simultaneously swap $w_{1}\leftrightarrow w_{2},v_{1}\leftrightarrow v_{2}$ then the original system is recovered. Note that $H$ only has inversion symmetry if $w_{1}=w_{2},v_{1}=v_{2}$. This condition (\ref{eq:inv}) tells us that the eigenvectors obey $\left|\psi\left(k,w_{1},w_{2},v_{1},v_{2}\right)\right\rangle =\sigma_{x}\left|\psi\left(-k,w_{2},w_{1},v_{2},v_{1}\right)\right\rangle $. In order to keep the forthcoming notation succinct we will write this as $\left|\psi\left(k,1\right)\right\rangle =\sigma_{x}\left|\psi\left(-k,2\right)\right\rangle $. Explicitly
\begin{equation}
\left|\psi_{k,1}\right\rangle =\left(\begin{array}{c}
e^{-i\phi_{k,1}}\cos\theta_{k,1}\\
\sin\theta_{k,1}
\end{array}\right)
,\left|\psi_{-k,2}\right\rangle =\left(\begin{array}{c}
e^{i\phi_{k,1}}\sin\theta_{k,1}\\
\cos\theta_{k,1}
\end{array}\right).
\end{equation}
Note that $\phi_{k,1}=-\phi_{-k,2}$ and $\sin\theta_{k,1}=\cos\theta_{-k,2}$. We wish to show that the real Berry phase $Q^{r}=i\int_{BZ}\left\langle \psi_{k}\right|\frac{\partial}{\partial k}\left|\psi_{k}\right\rangle dk$ is only quantized if inversion symmetry is obeyed. Consider the following
quantity
\begin{equation*}
Q_{1}^{r}\pm Q_{2}^{r}=\int_{BZ}\dot{\phi}_{k,1}\cos^{2}\theta_{k,1}dk\pm\int_{BZ}\dot{\phi}_{k,2}\cos^{2}\theta_{k,2}dk.
\end{equation*}
If $Q_{1}^{r}$ and $Q_{2}^{r}$ are quantized individually then so will their sum and difference. We break this quantity up as
\begin{eqnarray*}
Q_{1}^{r}\pm Q_{2}^{r} & &= \int_{0}^{\pi}\dot{\phi}_{k,1}\cos^{2}\theta_{k,1}dk+\int_{-\pi}^{0}\dot{\phi}_{k,1}\cos^{2}\theta_{k,1}dk\label{eq:sumDiff}\\
 &  & \pm \left( \int_{0}^{\pi}\dot{\phi}_{k,2}\cos^{2}\theta_{k,2}dk+\int_{-\pi}^{0}\dot{\phi}_{k,2}\cos^{2}\theta_{k,2}dk \right) \nonumber .
\end{eqnarray*}
In order to evaluate $Q_{1}^{r}+Q_{2}^{r}$, consider the sum of the first and last integral in the expression above
\begin{eqnarray*}
 &  & \int_{0}^{\pi}\dot{\phi}_{k,1}\cos^{2}\theta_{k,1}dk+\int_{-\pi}^{0}\dot{\phi}_{k,2}\cos^{2}\theta_{k,2}dk\\
 & = & \int_{0}^{\pi}\dot{\phi}_{k,1}\cos^{2}\theta_{k,1}dk+\int_{-\pi}^{0}\dot{\phi}_{k,2}dk-\int_{-\pi}^{0}\dot{\phi}_{k,2}\sin^{2}\theta_{k,2}dk\\
 & = & \int_{0}^{\pi}\dot{\phi}_{k,1}\cos^{2}\theta_{k,1}dk+\int_{-\pi}^{0}\dot{\phi}_{k,2}dk-\int_{0}^{\pi}\dot{\phi}_{k,1}\sin^{2}\theta_{-k,2}dk\\
 & = & \int_{-\pi}^{0}\dot{\phi}_{k,2}dk.
\end{eqnarray*}
Repeating this calculation with the second and third terms amounts to the same result with the replacement $1\leftrightarrow2$
\begin{eqnarray}
Q_{1}^{r}+Q_{2}^{r} & = & \int_{-\pi}^{0}\dot{\phi}_{k,2}dk+\int_{-\pi}^{0}\dot{\phi}_{k,1}dk\\
 & = & \int_{-\pi}^{\pi}\dot{\phi}_{k,2}dk
\end{eqnarray}
which is clearly an integer multiple of $\pi$.

In the inversion symmetric case $Q_{1}^{r}=Q_{2}^{r}$ hence it is clear that $Q_{1}^{r}-Q_{2}^{r}=0$. Thus we have just demonstrated that if the system possesses inversion symmetry then its real Berry phase will be an integer multiple of $\pi$.

Without assuming inversion symmetry, note $Q_{1}^{r}-Q_{2}^{r}=\left(Q_{1}^{r}+Q_{2}^{r}\right)-2Q_{2}^{r}$ which allows us to make use of the expression which we just derived
 \begin{equation}
Q_{1}^{r}-Q_{2}^{r}=\int_{-\pi}^{\pi}\dot{\phi}_{k,2}dk-2\int_{-\pi}^{\pi}\dot{\phi}_{k,2}\cos^{2}\theta_{k,2}dk. \label{eq:phaseDiff}
\end{equation} We use integration by parts on the second integral, then rewrite it in a more symmetric form as
\begin{widetext}
\begin{equation}
Q_{1}^{r}-Q_{2}^{r}=\left.\phi_{k,2}\left(1-2\cos^{2}\theta_{k,2}\right)\right|_{-\pi}^{\pi}+2\left(\int_{-\pi}^{\pi}\phi_{k,2}\sin\theta_{k,2}\cos\theta_{k,2}\dot{\theta}_{k,2}dk-\int_{-\pi}^{\pi}\phi_{k,1}\sin\theta_{k,1}\cos\theta_{k,1}\dot{\theta}_{k,1}dk\right).
\end{equation}
\end{widetext}
In the inversion symmetric case both terms independently go to zero. In the absence of inversion symmetry,  we do not expect this expression to be quantized in general.

\section{\label{sec:apB} Inversion-symmetric, chiral model}

We discuss aspects of the non-Hermitian, chiral, SSH model since such a Hamiltonian  serves as the simplest starting point to understand the role of non-Hermiticity in topological models. Furthermore, we will restrict our analysis to inversion-symmetric models since these are relevant to plasmonic experiments mentioned in the main text.

Begin by parameterizing the bulk Hamiltonian (\ref{eq:bulkH}) according to $v_{1}=v_{2}=e^{i\chi_{v}}\sin t,w_{1}=w_{2}=e^{i\chi_{w}}\cos t,u=0$.
This is the most general chiral, inversion-symmetric SSH model. The bulk dispersion reads
\begin{equation}
E_k^2 = e^{2i\chi_w} \cos^2 t + e^{2i\chi_v} \sin^2 t + 2 e^{i(\chi_v+\chi_w)}\cos t \sin t \cos k
\end{equation}
It is clear that in order to obtain a band closing in the bulk spectrum, the parameter $t$ must be tuned to the value $t=(2n+1)\pi/4,n\in\mathbb{Z}$. Thus the lack of Hermiticity does not change the topological transition point (which occurs at a band closing) in the inversion-symmetric model. The same is true if inversion symmetry is broken by just a phase term in the hopping elements. It is also worth noting that by tuning the phase degrees of freedom $\chi_{v,w}$ we can choose the momentum
$k_{\text{gap}}$ at which the gap closes in the bulk spectrum according to $\pm k_{\text{gap}} =\chi_w -\chi_v \pm \pi $, where relevant $k$ values occur in the first Brillouin zone.

If we expand around the topological transition point, it is possible to determine how the magnitude of the energy gap $\Delta$ scales as a function of $t$ away from the transition point.  Keeping leading orders in $ \delta t$ leads to the following expression
\begin{equation}
\Delta(\delta t)^4 = \delta t^2\left[1-\cos(2(\chi_w-\chi_v)) + O(\delta t^2) \right].
\end{equation}
The non-Hermitian, chiral model shows a square-root scaling behavior $\Delta(\delta t) \propto \sqrt{\delta t}$ in the magnitude of the energy (unless $\chi_w-\chi_v=n\pi,n \in \mathbb{Z}$). This is to be contrasted with the linear scaling of the Hermitian SSH model. Indeed this is confirmed numerically in Fig. \ref{fig:appen}(a).

\begin{figure}
\begin{centering}
\includegraphics[scale=0.1]{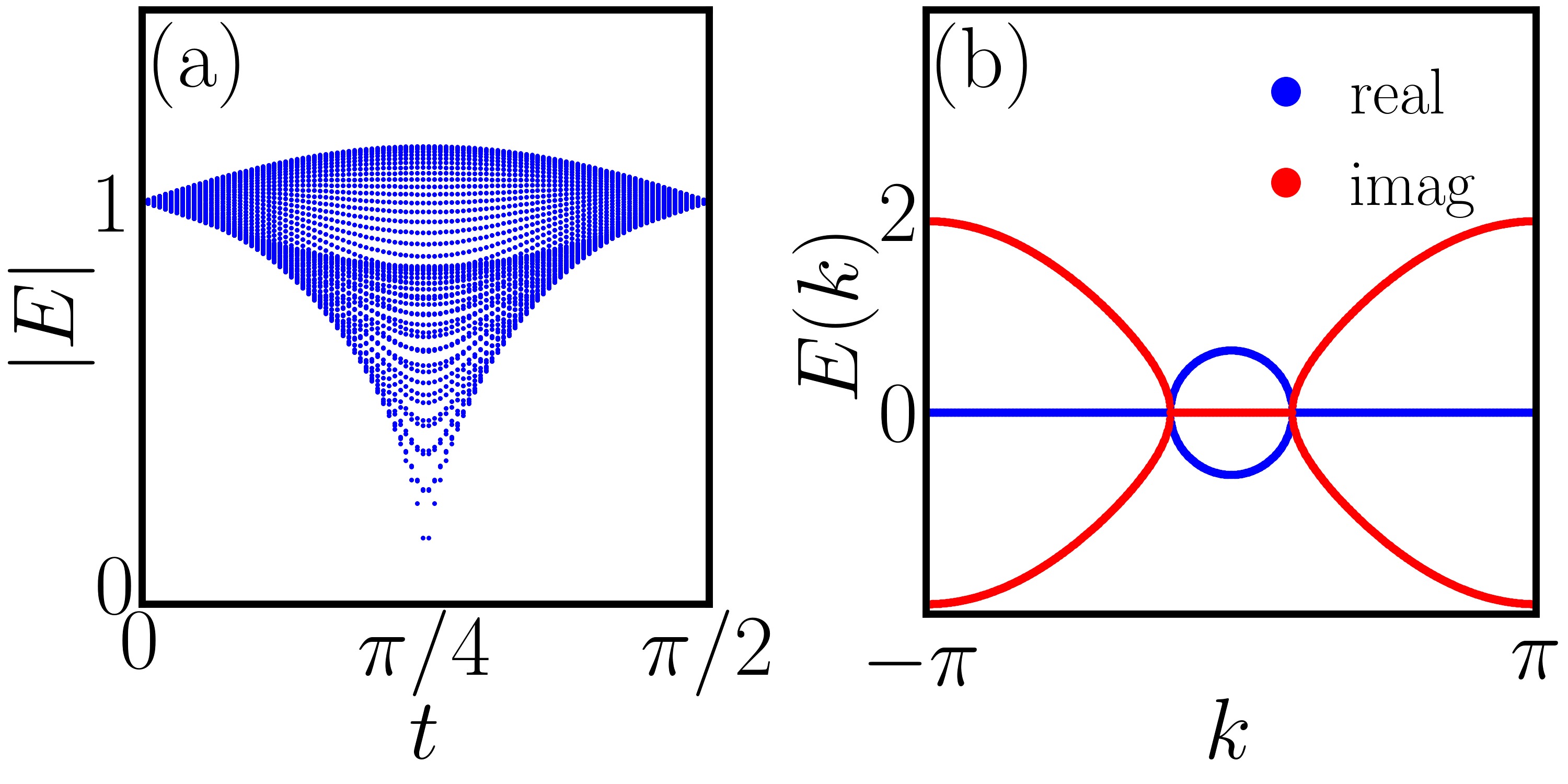}
\par\end{centering}
\caption{\label{fig:appen} \label{fig:appen}(a) Magnitude of finite chain spectrum with periodic boundary conditions (no edge modes) for $v=\exp i\chi_v \sin t$, $w=\exp i\chi_w \cos t$, $\chi_v=\pi/5,\chi_w=0$. The magnitude of the energy gap scales as $\sqrt{t}$ away from the topological transition. (b) Bulk dispersion at transition point $t=\pi/4$ when $\chi_v=-\chi_w=0.3(2\pi)$. The spectrum at any $k$ is either purely real or imaginary.}
\end{figure}

In the special case when $\chi_v=-\chi_w=\chi$, then there is an emergent, anti-linear symmetry at the topological phase boundary which ensures that the energetic secular equation will be real. Specifically, for the bulk Hamiltonian (\ref{eq:bulkH}) at the transition point $H(\chi,k)=H(-\chi,-k)^*$. The sign of $\chi$ cannot effect the bulk spectrum due to periodic boundary conditions. Therefore this relation constitutes an anti-linear symmetry which implies that eigenvectors are either broken or unbroken, in analogy with the $PT$-symmetric case. The combination of the chiral and anti-linear symmetry implies that energies are either purely real or imaginary. This can be seen in Fig. \ref{fig:appen}(b), where the spectrum undergoes a broken-to-unbroken transition in $k$.

We have refrained from commenting on the chiral, inversion-broken case in the event of unequal magnitudes, \textit{i.e.} $\left|v_{1}\right|\neq\left|v_{2}\right|,\left|w_{1}\right|\neq\left|w_{2}\right|$. This is known as the Hatano-Nelson model, and has recently been proposed in optical lattices \cite{Hatano1,Longhi2,Longhi3,Longhi4}. While our cBerry phase analysis shows that this model is expected to undergo a topological transition, it is difficult to numerically confirm this assertion. In the case of non-periodic boundary conditions, spectrum convergence with respect to system size is slow.

Finally, we mention that if additional degrees of freedom are added to the model (such as beyond nearest-neighbor hopping), then it is possible to achieve any integer invariant $\mathbb{Z}$. These systems possess more  than one pair of edge modes. This confirms the $\mathbb{Z}$ classification of the model.

\bibliography{sshBib} 
\bibliographystyle{apsrev4-1}

\end{document}